# Using AI-Generated Feedback to Improve Critical Thinking and Writing Proficiency

Qi Zhu[1], Xiaoming Zhai[2][0000-0003-4519-1931], Yan Zou[1], Chunlei Gao[1][0009-0009-7006-293X]

[1] School of Education, Jiangxi Normal University, 99 Ziyang Ave., Nanchang 330022, Jiangxi, China

[2] University of Georgia, Athens, GA 30666, USA

Zhuqi77@jxnu.edu.cn

Xiaoming.Zhai@uga.edu

Zouyan2024@jxnu.edu.cn

gaochunlei@jxnu.edu.cn

**Abstract.** Research indicates students require customized written composition feedback to enhance critical thinking and writing competence, yet teachers face barriers to delivering timely personalized guidance due to heavy workloads. To address this gap, this study developed the Writing Improvement and Smart Evaluation Agent (WISE Agent), an artificial intelligence (AI) feedback tool targeting textual logic and perspective biases in student essays. We conducted a three-month intervention with 260 Chinese sixth-grade students, each completing seven themed essays and receiving targeted WISE Agent feedback shortly after submission. Assessment used a critical thinking rubric adapted from the California Critical Thinking Disposition Inventory (CCTDI), covering seven core dimensions including cognitive maturity and open-mindedness. Results indicate structural optimizations in critical thinking dimensions rather than a uniform increase in total scores. While the overall enhancement was non-significant, granular analysis revealed divergent developmental trajectories: lower-performing students advanced in evidence selection, while higher-per- forming students excelled in perspective diversification. This study suggests that WISE Agent serves as an effective cognitive scaffold, facilitating incremental critical thinking development through personalized feedback loops, providing a practical pathway for AI-supported critical thinking cultivation.

**Keywords:** Artificial Intelligence (AI), Automated Feedback, Elementary School Chinese, Essay Writing, Critical Thinking.

## 1 Introduction

In the era of global digital transformation, fostering higher-order thinking has become a core objective for basic education [11]. Writing, as a critical cognitive process reconstructing implicit thought into explicit logic, provides an essential pathway to develop logical reasoning and dialectical analysis skills [3]. However, constrained by heavy grading burdens and experiential disparities, traditional writing feedback is often lagged and confined to surface-level evaluations, failing to trigger deep cognitive engagement [9]. Breakthroughs in Generative Artificial Intelligence (GenAI) have opened possibilities for constructing dynamic, scalable cognitive scaffolds [7], yet mechanisms for promoting higher-order thinking remain understudied. Current research often positions AI as a linguistic accuracy editor, overlooks elementary students in critical cognitive transitions, and lacks clarity on differentiated impact mechanisms across baseline proficiency levels [5]. Adopting a mixed-methods approach, this study deployed the Writing Improvement and Smart Evaluation Agent (WISE Agent) in a three-month sixth-grade Chinese language teaching intervention. Integrating longitudinal quantitative tracking and qualitative analysis, this study addresses two core research questions: (1) What is the intervention effect of intelligent feedback on students' critical thinking? (2) Under the intervention, what differential evolutionary trajectories do writing performance and critical thinking development exhibit among students with different academic proficiencies?

## 2 Related Work

Automated Writing Evaluation (AWE) has evolved from rudimentary grammar- checking tools into sophisticated systems capable of multi-dimensional linguistic analysis [17]. Early research primarily emphasized the efficiency of these tools in providing immediate, surface-level feedback to alleviate instructors' administrative burden [17]. However, as pedagogical objectives shift toward fostering higher-order thinking, traditional automated feedback has been criticized for its reductionist approach, which prioritizes syntactic accuracy at the expense of semantic coherence and argumentative depth [4].

Modern academic perspectives suggest that Artificial Intelligence (AI) driven feedback mechanisms must transcend automated scoring to facilitate the "writing-to- learn" paradigm, focusing on evaluating complex constructs such as argumentative quality and evidence integration [15]. Recent empirical studies demonstrate that automated feedback embedded with AI-generated metacognitive prompts can effectively guide learners to scrutinize evidence validity and optimize argumentative logic by integrating multiple perspectives [12]. Notably, the efficacy of AI feedback is moderated by learners' prior knowledge, resulting in distinct differentiated interaction patterns across proficiency levels [14]. High-proficiency learners tend to utilize intelligent suggestions as catalysts for deep dialectical refinement, whereas their less-proficient peers rely more heavily on AI-provided logical frameworks for structural knowledge integration [2].

Despite these advances, critical gaps remain in understanding how to design adaptive AI feedback systems for elementary students in a cognitive transition phase. Existing research focuses primarily on secondary and higher education contexts, with limited empirical evidence on how to balance linguistic precision with cognitive depth in primary education settings. This study addresses this gap by investigating how AI feed-back can support critical thinking development through structured writing interventions, drawing on distributed cognition theory and writing-to-learn pedagogical frameworks to inform system design and evaluation.

# 3 Research Methods

## 3.1 Participants

The study involved 260 Chinese sixth-grade students (aged 11–13) from a public primary school. Integrated into a three-month Chinese language arts curriculum, this longitudinal intervention ensures high ecological validity within a naturalistic classroom setting. Following strict ethical protocols regarding informed consent and anonymization, participants were stratified based on pre-test scores from a validated critical thinking disposition scale (Cronbach's $\alpha = 0.935$, KMO = 0.883). Using an optimal discrimination index, students were categorized into three proficiency cohorts for granular analysis: the High-Level Group (27%, n=70), the Middle-Level Group (46%, n=119), and the Low-Level Group (27%, n=71).

## 3.2 Intervention Design

This study employed a standardized, curriculum-aligned intervention framework integrating writing task design, multidimensional evaluation and AI-supported feedback [11]. Tasks were sourced from the national Grade 6 Chinese textbook, covering seven units across narration, practical, and imaginative writing [1]. A two-dimensional evaluation framework—encompassing writing quality (organization, objectivity, and argumentation depth) and critical thinking (logical reasoning, multiple perspectives, and dialectical analysis)—underpinned the feedback generation. This framework was developed through collective teacher consultation and adapted from the CCTDI scale to suit elementary cognitive characteristics [13].

The intervention was implemented via WISE Agent, an AI tool built on the Coze platform and powered by the DeepSeek Large Language Model (LLM). The system follows a closed-loop workflow: (1) Essay Recognition: Handwritten manuscripts undergo standardized OCR and image-stitching to ensure integrity; (2) Feedback Generation: Acting as a "Senior Primary Teacher," the LLM provides qualitative, tier-based assessments (e.g., "Developing" to "Advanced") with actionable scaffolding, such as prompting sensory details or counterarguments; and (3) Iterative Revision: Students revise compositions independently before final teacher review. To ensure reliability, feedback adheres to predefined rubrics and standardized prompts refined through pilot testing (N=120). Security is maintained via content anonymization and end-to-end encryption. Achieving an average correction cycle of 40 seconds, the system significantly enhances feedback timeliness compared to traditional manual methods.

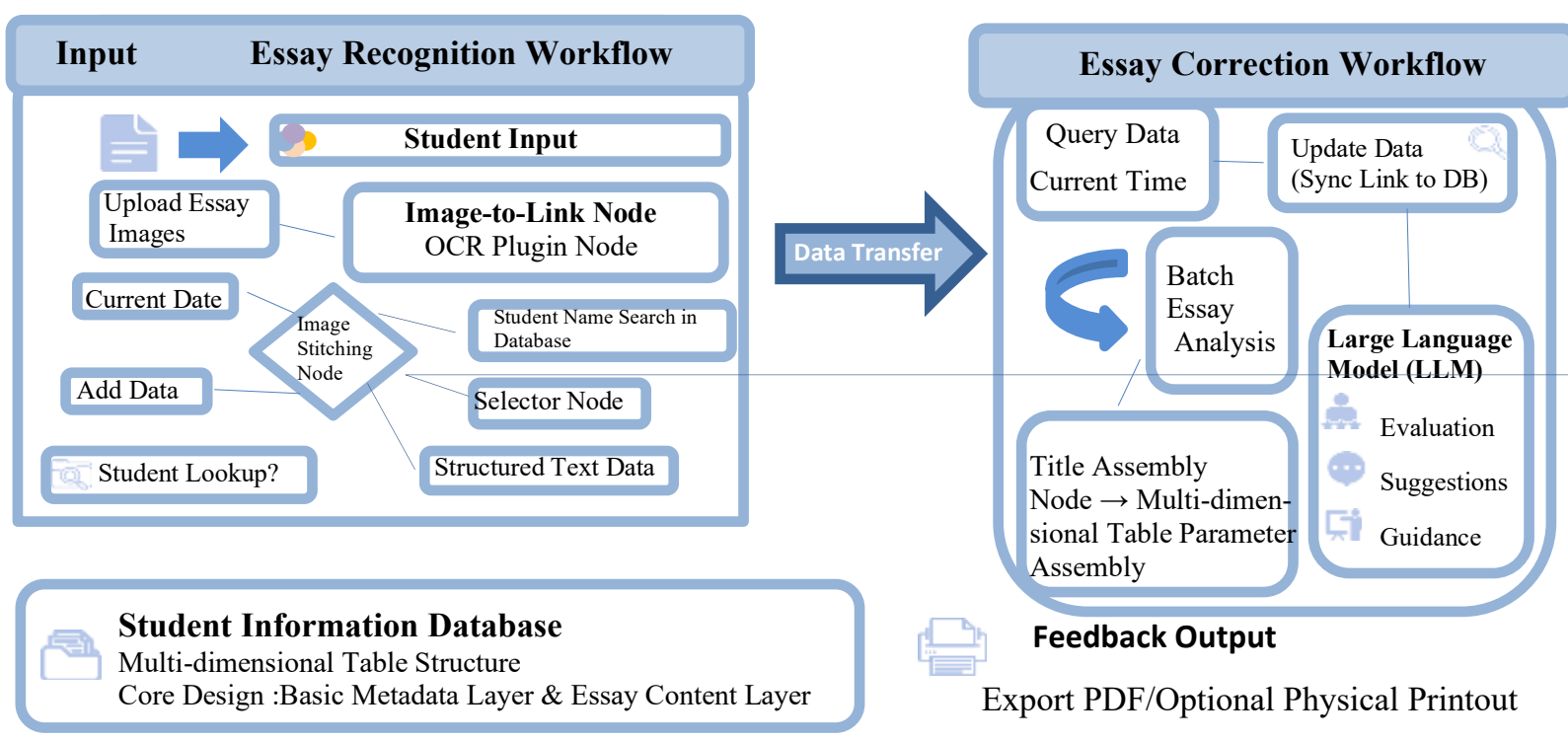

**Fig. 1.** Flowchart of the Intelligent Agent's Operation Mechanism

### 3.3 Measures

Following the psychometric foundation established in Section 3.1, this section operationalizes the evaluation criteria. To bridge the potential gap between cognitive "disposition" and writing "performance," we established a robust conceptual mapping between the CCTDI theoretical framework and a customized writing rubric. This framework was developed based on the Compulsory Education Curriculum Standards for Chinese (2022 Edition) and refined through two rounds of expert syn-thesis: a theoretical review by five pedagogy specialists to ensure structural alignment, followed by collaborative deliberations among 14 senior language teachers to ensure pedagogical feasibility.

As shown in Table 1, the core CCTDI dimensions were operationalized into specific writing performance indicators. This multi-layered validation ensures that the WISE Agent's feedback on writing logic not only aligns with national curriculum goals but also directly nourishes the underlying cognitive dispositions. The disposition scale thus serves as a measure of "potential" that is activated and refined through iterative writing tasks. Ultimately, this integrated measurement framework provides a standardized benchmark for tracking cognitive development trajectories, ensuring that the empirical analysis of the WISE Agent-mediated intervention remains both rigorous and methodologically consistent.

**Table 1.** Mapping between Core CCTDI Dispositions and Writing Rubric Indicators

| CCTDI Dispositions | Writing Performance Indicators |
|---|---|
| Open-mindedness | Perspective Diversification & Bias Identification |
| Truth-seeking | Evidence Sourcing & Argumentation Depth |
| Inquisitiveness | Intellectual Depth & Information Synthesis |
| Systematicity | Textual Organization & Logical Expression |

# 4 Results

## 4.1 Critical Thinking Disposition Changes

This study conducted paired-sample t-tests and effect size calculations using SPSS 26.0 to assess shifts in critical thinking dispositions. Pre-test scores confirmed the homogeneity of variance (p=0.32), satisfying the prerequisite assumptions for parametric testing. Results indicated that after the three-month WISE Agent intervention, the aggregate CT disposition scores improved, though the increment did not reach statistical significance (t=-0.854, p=0.394, Cohen's d=0.054). However, granular analysis at the sub-dimension level revealed significant structural reorganizations. Scores in Truth- seeking (t=-5.138, p<0.001) and CT Self-confidence (t=-3.038, p=0.003) exhibited significant upward trends. Conversely, the Cognitive Maturity dimension score experienced a significant decline (t=2.358, p=0.019). This specific de-crease, rather than indicating a regression in intellectual capacity, likely reflects a "de-centering" process. As students engaged with the multi-dimensional perspectives provided by the WISE Agent, their initial "naive optimism" was challenged, leading to a more sober recognition of cognitive complexity and a subsequent correction of previous overestimation. No significant changes were observed in Open-mindedness, Inquisitiveness, Analyticity, or Systematicity. In summary, while the intervention did not yield a linear leap in overall CT levels, it catalyzed a critical internal structural adjustment, characterized by differentiated evolutionary trends across sub-dimensions.

## 4.2 Longitudinal Trajectories and Heterogeneous Develop-mental Characteristics

Combining longitudinal quantitative tracking and thematic analysis methods, this study conducted a comparative analysis of the developmental trajectories of students with different academic proficiency levels. Data showed significant differences in the improvement magnitude of critical thinking among different groups: high-level students' critical thinking scores increased by 18%, low-level students by 8%, and intermediate-level students by 12%. The intermediate group exhibited a growth trajectory (12%) that bridged the gap between high- and low-proficiency cohorts. In terms of developmental trajectories, the high-level group showed characteristics of high-level stability, with coefficients of variation for writing and thinking scores maintained within a low fluctuation range of 0.02~0.04 throughout the intervention cycle; the low-level group showed obvious stage-based fluctuation characteristics, with a time-lag effect where improvement in thinking ability preceded improvement in writing performance. The mechanism of intelligent feedback presented asymmetric adaptation characteristics: for low-level students, it focused on providing structured writing scaffolds to help them improve the standardization of evidence selection and rigor of logical argumentation; for high-level students, it adopted counter-questioning heuristic strategies to guide them to expand multiple perspectives and dialectical analysis capabilities, ultimately achieving differentiated development goals for different groups.

## 5 Discussion

This study demonstrates that the WISE Agent facilitates the synergistic development of critical thinking (CT) and writing through cognitive restructuring rather than linear growth. Although aggregate scores remained stable, significant structural reorganization occurred: gains in Truth-seeking and Self-confidence were offset by a decline in Cognitive Maturity. These asymmetrical trajectories—particularly the accelerated dialectical analysis in high-performing students—confirm that improvements were catalyzed by the agent's adaptive feedback rather than general maturation.

The decline in Cognitive Maturity aligns with the Dunning-Kruger effect [8]; multidimensional feedback corrected students' initial epistemic overestimation, transitioning them from a self-centered stance toward recognizing cognitive complexity. This "intellectual recalibration" is a necessary precursor to higher-level schema restructuring [6]. However, the observed "time-lag effect" in lower-performing cohorts suggests that high-density feedback may induce cognitive overload. Future iterations should refine feedback granularity to prevent "over-scaffolding."

Practically, this research underscores the potential of human-machine collaboration. To mitigate the Matthew Effect, future models should integrate AI-driven scaffolding with teacher-led "offloading." While curriculum constraints precluded a randomized control group, the study's high ecological validity provides a robust foundation. Subsequent research should employ quasi-experimental designs to validate the long-term scalability of such personalized interventions.

## 6 Conclusion

This three-month study examined the WISE Agent's impact on primary students' critical thinking (CT) and writing. While the intervention yielded no statistically significant leap in aggregate CCTDI scores, it triggered meaningful structural optimizations in key sub-dimensions. Notably, significant growth in Truth-seeking and CT Confidence indicates a pivot from passive knowledge-telling toward active inquiry. The lack of overall significance likely stems from the short duration required to internalize complex cognitive habits; however, these localized gains confirm the tool's value as a specialized intervention for specific cognitive bottle-necks. Rather than a mere corrective editor, the WISE Agent functions as a formative heuristic mediator, translating abstract CT requirements into actionable writing strategies. The observed differentiated response patterns across proficiency cohorts further demonstrate that adaptive AI feedback can mitigate the "one-size-fits-all" limitation of traditional instruction. This study provides a practical pathway for AI-supported CT cultivation and an empirical reference for the digital transformation of primary language education.

**Disclosure of Interests.** The authors have no competing interests to declare.